\documentclass[pre,aps,preprint,amsmath,superscriptaddress]{revtex4}

\usepackage{graphicx}
\usepackage{dcolumn}
\usepackage{amsmath}
\usepackage{txfonts}
\usepackage{bm}

\usepackage[colorlinks,
            linkcolor=blue,       
            anchorcolor=blue,  
            citecolor=blue,        
            ]{hyperref}
 \usepackage{siunitx} 
 \usepackage{chemfig} 
 \usepackage{setspace}

\begin{document}
\begin{center}
{\large{\bf Click Metamaterials: Fast Acquisition of Thermal Conductivity and Functionality Diversities}}
\\ \hspace*{\fill} \\
Chengmeng Wang,\textsuperscript{1, *} Peng Jin,\textsuperscript{1, *, †} Fubao Yang,\textsuperscript{1} Liujun Xu\textsuperscript{2, ‡} and Jiping Huang\textsuperscript{1, §}\\
\textsuperscript{1}Department of Physics, State Key Laboratory of Surface Physics, and Key Laboratory of Micro and Nano Photonic Structures (MOE), Fudan University, Shanghai 200438, China\\
\textsuperscript{2}Graduate School of China Academy of Engineering Physics, Beijing 100193, China\\
\textsuperscript{†}Electronic address: 19110190022@fudan.edu.cn\\
\textsuperscript{‡}Electronic address: ljxu@gscaep.ac.cn\\
\textsuperscript{§}Electronic address: jphuang@fudan.edu.cn\\
\textsuperscript{*}These authors contributed equally to this work.
\end{center}

\date{\today}

\begin{center}
\large{Abstract}
\end{center}

Material science is an important foundation of modern society development, covering significant areas like chemosynthesis and metamaterials. Click chemistry provides a simple and efficient paradigm for achieving molecular diversity by incorporating modified building blocks into compounds. In contrast, most metamaterial designs are still case by case due to lacking a fundamental mechanism for achieving reconfigurable thermal conductivities, largely hindering design flexibility and functional diversity. Here, we propose a universal concept of click metamaterials for fast realizing various thermal conductivities and functionalities. Tunable hollow-filled unit cells are constructed to mimic the modified building blocks in click chemistry. Different hollow-filled arrays can generate convertible thermal conductivities from isotropy to anisotropy, allowing click metamaterials to exhibit adaptive thermal functionalities. The straightforward structures enable full-parameter regulation and simplify engineering preparation, making click metamaterials a promising candidate for practical use in other diffusion and wave systems.


\noindent{\textbf{Teaser:} Borrowing brilliance from click chemistry, click metamaterials offer a leap forward in material science, using versatile hollow-filled units to master thermal behaviors, paving the way for design simplicity and various functional applications.}

\section{\label{sec:level1}Introduction}

Heat energy is the most fundamental energy form, whose widespread existence and huge quantity make heat control particularly crucial for human society. Thermal metamaterials have contributed to uncovering unexpected heat transfer mechanisms due to their extraordinary abilities to manipulate heat flow \cite{PR_yangshuai,nrp_Zhang,convertive_AM_liying,ilusionzhu15,Trans_multithermotics_xu20}, such as nonreciprocal transmission \cite{convertive_AM_liying,Nonreciprocity_apl_xu}, topological transport \cite{Antiparity_liying,EP_Xu_pnas}, non-Hermitian physics \cite{topology_xu_prl,Diffusive_topology_Xu,Geometric_phase_xu21} and other unique thermal mechanisms \cite{Ultra_tc_qiu,qipan,thermal_emission_nc}. Various practical applications have also been proposed based on thermal metamaterials \cite{PR_yangshuai,convergent_transfer_shen}, such as cloaking \cite{APL_Fan,bilayer,3D_bilayer,First_Experiment_cloak_2012_Guenneau,transparency_xu21,jap_dai_18,omnithermotics_xu20,active_cloak}, concentrating \cite{First_Experiment_cloak_2012_Guenneau,concentrator_yu,porous_media_yfb}, rotating\cite{first_rotator_Narayana,second_rotator_Guenneau}, and camouflaging \cite{chameleonlike_rotator_Yangubao,chameleonlike_metashell_cloak_Xu,Chameleonlike_miscrofludics_cloak_Xu}. Exisitng thermal metamaterial designs are often based on transformation thermotics \cite{APL_Fan,Jin2022TunableLH,First_Experiment_cloak_2012_Guenneau,heterogeneously_architected_structure_xu}, scattering cancellation \cite{bilayer,3D_bilayer}, numerical optimization \cite{TFCs_Qiu,Jin2021ParticleSO,Topologyoptimized_1,Topologyoptimized_2}, or spatiotemporal modulation \cite{ThermalWC_xlj_prl,spatiotemporal_modualtion_yang,DiffusiveFD_xu_prl}. However, due to inherent barriers, these fragmented approaches usually make thermal metamaterial designs case by case, reducing design flexibility and functional diversity \cite{qipan}. On the one hand, common metamaterials have fixed thermal conductivities and necessitate the fabrication of an entirely new device when the target thermal conductivity changes \cite{TFCs_Qiu,TFC_NPJ}, making the original device useless and causing resource waste. On the other hand, though the effective thermal conductivity of rotating materials can be continuously tuned \cite{YingLi_KNI,EMT_Rotating_Qiu,Xu_Tunable,Jin2022TunableLH,ml_JP}, the method requires complex thermal convection, and it is hard to generate desired thermal anisotropy. Therefore, there is a pressing need to obtain convertible thermal conductivities using natural materials, which could offer a general platform for designing various thermal metamaterials and functionalities.
	
Click chemistry~\cite{Click_sharpless} has become one of the most influential paradigms in chemosynthesis due to its simplicity, flexibility, and reliability, which typically generates new molecules or materials by combining modified building blocks \cite{Click_book,click_chem_1,click_chem_2}. As a natural association, click chemistry may inspire thermal metamaterial designs to solve the abovementioned problems, but several critical challenges still hinder this cutting-edge idea from being applied in advanced heat control. Firstly, lacking a thermal counterpart of the modified building blocks in click chemistry is fatal due to the absence of a fundamental mechanism for achieving convertible thermal conductivities. Secondly, how to use a limited number of basic units to form a variety of thermal functionalities remains elusive. Therefore, resorting to the idea of click chemistry for solving heat transfer problems is still challenging.

Here, we address these key issues and present the design paradigm of click metamaterials for fast realizing thermal conductivity and functionality diversities. We fabricate a click metashell, consisting of tunable hollow-filled cells (THFCs) which are the counterparts of the modified building blocks in click chemistry. It is important to note that the THFCs are not cut from the metashell but are integral to its structure, as depicted in Figure \ref{fig.1}(a). By delicately arranging these THFCs, we can achieve convertible thermal conductivities from isotropy to anisotropy. Combining neighboring THFCs with various thermal conductivities (e.g., a 2 $\times$ 2 THFC array named the expanded THFC) can conveniently adjust anisotropic thermal conductivities without refabrication. The highly tunable thermal conductivities of click metamaterials also yield adaptive thermal functionalities, shown in Figure \ref{fig.1}(d). We fabricate an adaptive thermal cloak composed of this click metashell, performing adaptive responses to variable background components. Besides environmental adaptation, we illustrate that this click metashell can realize switchable functionalities simply by modifying the structure of the expanded THFCs. The easy-to-prepare click metamaterials may also have implications for applications in optics \cite{optics_adaptive_1,optics_adaptive_2,optical_materials}, acoustics \cite{acoustics_adaptive}, chemistry \cite{chameleonlike_mass_diffusion}, electromagnetism \cite{EM_adaptive,EM_chameleonlike_cloak,Deeplearning_adaptive_microwave,ferrofluids_gao,jap_dong,pre_hjp}, and other fields \cite{stock_markets_08,classical_music,nanochannels}. 

\section{Results}

{\bf Convertible Thermal Conductivity.}

We introduce a design paradigm for click metamaterials, founded on the assessment of the effective thermal conductivity (ETC) of THFC. Initially, we determine the ETC of THFC under varying air fractions, as elaborated in section \ref{sec:level2.1}, and subsequently propose a design strategy that facilitates convertible thermal isotropy. Following this, we calculate the properties of the expanded THFC, leveraging the ETC obtained from a single THFC, as detailed in section \ref{sec:level2.2}. This calculation enables us to attain convertible thermal conductivities ranging from isotropy to anisotropy, thereby allowing flexible manipulation of thermal anisotropy. The insights gleaned from this section lay the theoretical groundwork for the fabrication of metadevices with adaptive thermal functionalities.

\subsection{Convertible Thermal Isotropy}\label{sec:level2.1}

\begin{figure}
		\centering
		\includegraphics[width=0.8\linewidth]{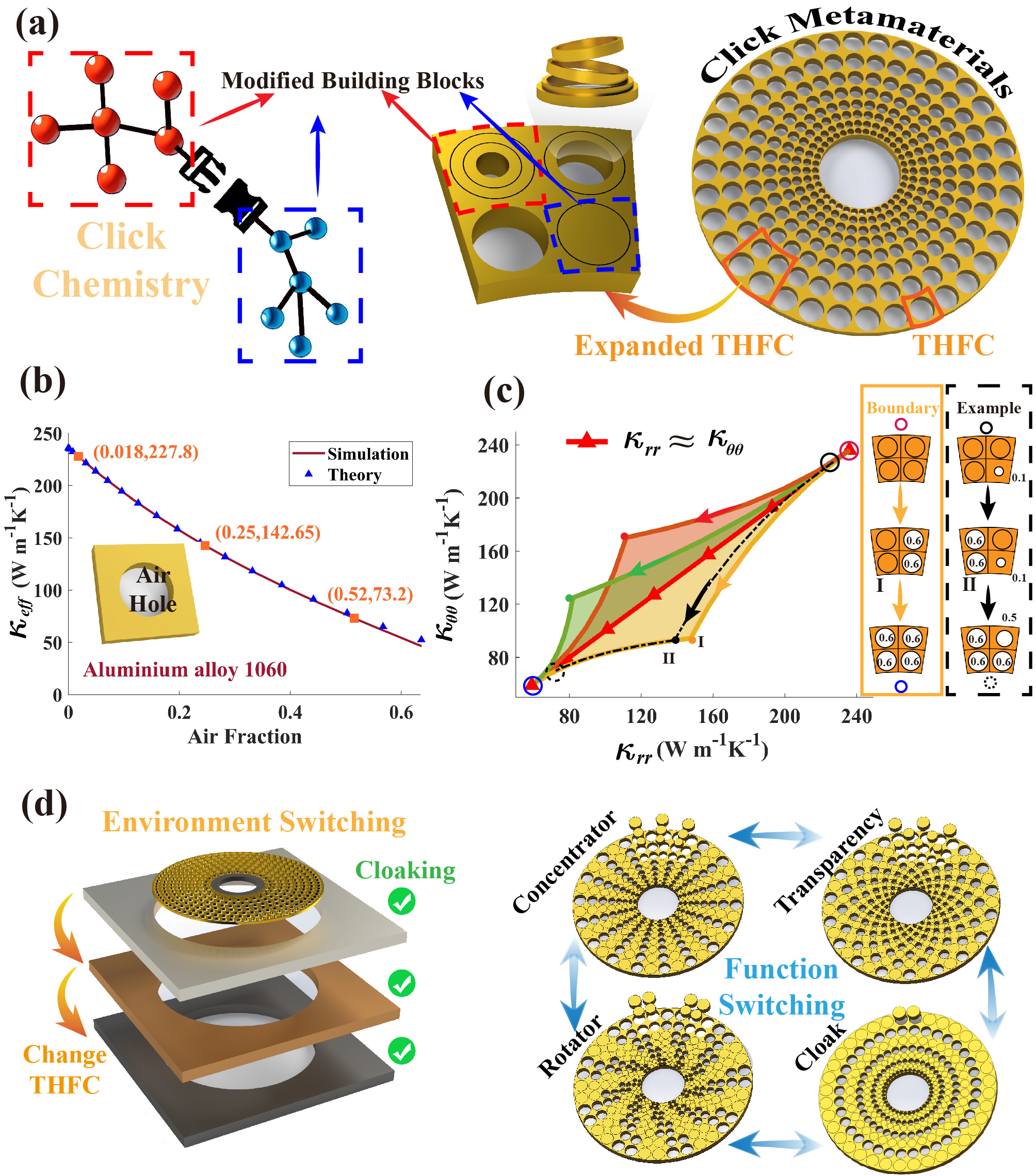}
        \caption{\fontsize{10.5bp}{17bp}(a): The relationship between click chemistry and click metamaterials. Click metashell is divided into a series of tunable hollow-filled cells (THFCs), constructed to mimic the modified building blocks in click chemistry. Click metashell with each THFC staying in the "hollow" state is depicted on the right. We use several concentric annuli to adjust the THFC's air fraction, whose state can be switched between "hollow" and "filled". The neighboring four THFCs combine the expanded THFC. (b): Physical foundation of the click metamaterials within the THFC. The THFC can be considered as a quasi-square flake. Correspondingly, the effective thermal conductivity of a square Aluminum alloy 1060 (236 W m$^{-1}$ K$^{-1}$) with a round air hole (0.02 W m$^{-1}$ K$^{-1}$) is also depicted. The solid blue triangles are the calculated results. And the solid red line is the simulation result. (c): Physical foundation of the click metamaterials within the expanded THFC. The anisotropic space when changing air fractions of the expanded THFC (the host medium is also Aluminum alloy 1060). Different air hole changing methods allow thermal conductivity to travel in different paths and directions. The orange region: $\kappa_{rr} \textgreater \kappa_{\theta\theta}$; The red and green region: $\kappa_{rr} \textless \kappa_{\theta\theta}$; The connecting line between the red triangles: $\kappa_{rr} \approx \kappa_{\theta\theta}$. (d): Illustrative applications of the click metamaterials. The click-reaction-induced thermal metadevices yield adaptive thermal functionalities to surroundings or application requirements.}
		\label{fig.1}
	\end{figure}

For adjusting the ETC of natural materials artificially, researchers often employ the effective medium theory \cite{EMT_Yang}. As mentioned above, we grid the metashell into tunable hollow-filled cells (THFCs), as shown in Figure \ref{fig.1}(a). We expect to control over the ETC of the metashell by modifying the characteristics of THFCs \cite{EMT_Jin}. Every THFC assembling the metashell is a quasi-square flake with a hollow structure. One can readily calculate the effective thermal conductivity ($\kappa_i$) of a square flake where an air hole locates at the center of it [see the model schematically illustrated in Figure \ref{fig.1}(b)] by directly solving the Laplace’s equations \cite{EMT_Yang}:
	
	\begin{equation}
		\kappa_i=\kappa_2\frac{[p+(1-p)L_i]\kappa_1+(1-p)(1-L_i)\kappa_2}{(1-p)L_i\kappa_1+[p+(1-p)(1-L_i)]\kappa_2}\label{eq1},
	\end{equation}
	where $\kappa_2$ is the thermal conductivity of the host medium. $p$ describes the air fraction of the host medium. $\kappa_1$ is the thermal conductivity of air. $L_i\,(i=a,b)$ (two-dimensional case) is the shape factor along the $x$-axis or $y$-axis, given by \cite{EMT_Yang}:
	
	\begin{equation}
		L_i=\frac{ab}{2}\int_{0}^{\infty}\frac{ds}{(i^2+s)\sqrt{(a^2+s)(b^2+s)}}\label{eq2}.
	\end{equation}
	In particular, we have $L_a=L_b=1/2$, supposing the air fraction is round. For a three-dimensional case, $L_c$ should be introduced as a third dimension and we should consider a spheriform air hole embedded in a cubic object, see Note S1 (Supplementary Materials) \cite{supplemental_material}.  The shape factor satisfies the summation rule, $\sum\limits_{i} L_i = 1$, $L_i \in [0,1]$.
	
As an example, we select a square piece of Aluminum alloy 1060 (236 W m$^{-1}$ K$^{-1}$) as the host medium. Then we obtain the effective thermal conductivity of the model (shown in Figure \ref{fig.1}(b)) with a round air hole (0.02 W m$^{-1}$ K$^{-1}$) via Eq.\ref{eq1}. The effective thermal conductivity of Aluminum alloy 1060 decreases from 236 to 46.3 W m$^{-1}$ K$^{-1}$ continuously as the air fraction increases from 0 to 0.6; see blue triangles in Figure \ref{fig.1}(b). To verify the analytical changing trends, we utilize the commercial software COMSOL Multiphysics to calculate the ETC of this model; see solid red line in Figure \ref{fig.1}(b). It's obvious that the theory and simulation fit well. In practice, we can resort to the map datasets between effective thermal conductivity and air proportion for the purpose of accurate adjustment. So we can equivalently control the effective conductivity of this model by modulating the radius of the round air hole. To exactly modify the radius, we can manufacture several concentric annuli with appropriate thicknesses (e.g. varying thicknesses with equal differences). Then, we put the annuli touched each other into the air hole to resize it, composing a THFC with targeted thermal conductivity, shown in Figure \ref{fig.1}(a).
	
Generally, the click reaction occurring in the procedure of realizing convertible thermal isotropy includes four steps. Firstly, according to the requirement of actual applications, we calculate the thermal conductivity of the metashell. Secondly, we grid the metamaterial region into a series of THFCs, with each THFC holding the calculated thermal conductivity. Thirdly, we consult the collected datasets to obtain the air hole sizes of the THFCs, corresponding to the calculated thermal conductivity. Finally, we adjust the amount of the concentric annulus of each THFC simultaneously, making the air hole with the calculated size. This methodology allows us to regulate the ETC of the metashell without refabricating an entirely new device.
	
\subsection{\label{sec:level2.2}Convertible Thermal Anisotropy}
	
We can also realize convertible thermal anisotropy of the metashell by the click reactions: the judicious combination of THFCs with different air fractions. We suppose the thermal conductivity of the metashell is a tensor matrix $\begin{bmatrix} 
		\kappa_{rr} & \kappa_{r\theta}\\\kappa_{\theta r} & \kappa_{\theta\theta} 
\end{bmatrix}$ ($\kappa_{r\theta} = \kappa_{\theta r}$) in the cylindrical coordinate system. For anisotropic ETCs, when $\kappa_{r\theta}=0$, we consider a 2 $\times$ 2 array comprising four neighboring THFCs with distinct air hole sizes to realize the convertible anisotropy of the metashell, which is dubbed expanded THFC hereinafter; see Figure \ref{fig.1}(a). The ETC of the upper left (mark as UL), upper right (mark as UR), lower left (mark as LL) and lower right (mark as LR) THFC in the expanded THFC are $\kappa_a$, $\kappa_b$, $\kappa_c$ and $\kappa_d$, respectively. Based on the ETCs of the four THFCs, we construct the analytical expression of the expanded THFC's ETC tensor (diagonal terms) using series and parallel models:
	
	\begin{subequations}
		\begin{equation}
			\kappa_{rr}=\frac{1}{2}\frac{\kappa_a\kappa_c}{a\kappa_a+b\kappa_c}+\frac{1}{2}\frac{\kappa_b\kappa_d}{a\kappa_b+b\kappa_d},
		\end{equation}
		\begin{equation}
			\kappa_{\theta\theta}=2a\frac{\kappa_c\kappa_d}{\kappa_c+\kappa_d}+2b\frac{\kappa_a\kappa_b}{\kappa_a+\kappa_b},
		\end{equation}
		\label{eq.3}
	\end{subequations}
where $a=1/(1+e^{2\delta\theta})$ and $b=1/(1+e^{-2\delta\theta})$. And we have ln$(r_{i+1}/r_i)=\delta\theta$ to ensure the THFC is quasi-square, where $\delta\theta$ is the central angle of a THFC phase and $r_i$ (i = 1,2,. . . ,n) is the inner radius of the $i$-th annular layer of the structure \cite{qipan}. When the air fraction of each THFC varies from 0 to 0.6, respectively, we obtain the $\kappa_{rr}$ and $\kappa_{\theta\theta}$ of the expanded THFC (Aluminum alloy 1060 is also the host medium) with different combinations of THFCs using Eq.\ref{eq.3}, as shown in Figure \ref{fig.1}(c). We label the quaternion (UL, UR, LL, LR) to represent the different combinations of air fractions within the expanded THFC and denote the corresponding ETC tensor of the expanded THFC as $\kappa_{\rm UL,UR,LL,LR}$. For example, $\kappa_{0,0,0,0} = \begin{bmatrix} 
	236 & 0\\0 & 236 
\end{bmatrix}$ where the quaternion is (0, 0, 0, 0) as denoted with the red solid circle in Figure \ref{fig.1}(c). $\kappa_{0.6,0.6,0.6,0.6} = \begin{bmatrix} 
		46.3 & 0\\0 & 46.3 
\end{bmatrix}$ where the quaternion is (0.6, 0.6, 0.6, 0.6) as denoted by the blue solid circle. Starting from (0, 0, 0, 0), we keep the air fraction of the UL and LL invariant and enlarge the counterpart of the UR and LR until the quaternion is (0, 0.6, 0, 0.6), the $\kappa_{rr}$ and $\kappa_{\theta\theta}$ walk directionally along the yellow boundary from the red solid circle to point \uppercase\expandafter{\romannumeral1}. Then we keep the air fraction of the UR and LR invariant and increase the counterpart of the UL and LL until the quaternion is (0.6, 0.6, 0.6, 0.6), the $\kappa_{rr}$ and $\kappa_{\theta\theta}$ continue to walk from point \uppercase\expandafter{\romannumeral1} to the blue solid circle. In addition, we can obtain the other two boundaries (red and green boundaries) by appropriate adjustment of quaternion; see Figure \ref{fig.1}(c). Furthermore, we obtain an anisotropic parameter space ($\kappa_{rr}$, $\kappa_{\theta\theta}$) considering different quaternions (UL, UR, LL, LR). For instance, when permitting three kinds of air fraction in the expanded THFC, the $\kappa_{rr}$ and $\kappa_{\theta\theta}$ can vary along the black dashed curve, commencing from the quaternion (0, 0, 0, 0.1). Thus, we can realize convertible thermal anisotropy within the expanded THFC.
	
Notably, the $\kappa_{rr}$ and $\kappa_{\theta\theta}$ of the expanded THFC depend only on $\delta\theta$ and the $\kappa$ values of its four individual THFCs. Therefore, the whole metashell can be regarded as a periodic arrangement of this 2 $\times$ 2 array, whose effective thermal conductivity is position-independent. Finally, we can realize convertible thermal anisotropy on the click metashell by setting the proper quaternion (UL, UR, LL, LR) of each expanded THFC, without loss of generality. For anisotropic ETCs, when $\kappa_{r\theta} \neq 0$, the ETC tensor $\begin{bmatrix} \kappa_{rr} & \kappa_{r\theta}\\\kappa_{\theta r} & \kappa_{\theta\theta} \end{bmatrix}$ can be realized by rotating the expanded THFCs (ETC is $\begin{bmatrix} \kappa_{rr} & 0 \\ 0 & \kappa_{\theta\theta} \end{bmatrix}$) with an angle $\theta$ \cite{TFC_NPJ}, see section \ref{sec:level3.2} in detail. Therefore, the ETC of the metashell covers full-parameter anisotropic space by the periodic arrangement of the expanded THFCs.
\\

{\bf Click-Metamaterials-Induced Adaptive Functionalities}. 

In this section, we delve into the practical applications of the click metamaterials. The attribute of convertible thermal isotropy enables us to construct an adaptive thermal cloak, constituted of the click metashell, which exhibits adaptive responses to fluctuating background components. Simultaneously, the property of convertible thermal anisotropy in the click metashell allows for a functional shift between various thermal states, including a thermal cloak, thermal concentrator, thermal transparency, and thermal rotator. Consequently, click metamaterials manifest adaptive thermal functionalities, demonstrating their significant potential in real-world applications.

\subsection{Convertible Isotropic Metashell With Functional Stability}\label{sec:level3.1}

To verify the convertible thermal isotropic conductivity of the metashell, we present an adaptive thermal cloak that incorporates this click metashell, allowing it to perform adaptive responses to variable background components. It is well-established that the bilayer thermal cloak can achieve perfect thermal cloaking with only bulk isotropic materials \cite{bilayer,3D_bilayer}. However, this invisible device only works in a specific background component, rendering it impractical when working scenes change. The bilayer cloak consists of an inner layer with thermal conductivity $\kappa_2$ and an outer layer with thermal conductivity $\kappa_3$, working in an environment with thermal conductivity $\kappa_0$. The inner layer is considered as a perfect insulation material. With the aim of achieving exact thermal cloaking, $\kappa_3$ and $\kappa_0$ should satisfy Eq.\ref{eq4a} in a three-dimensional (3D) case. Similarly, for a two-dimensional (2D) case, an analysis analogous to that for 3D has been proposed, which gives rise to the relationship stated in Eq.\ref{eq4b}.
	
	\begin{subequations}
		\begin{equation}
			\kappa_3 =\frac{2c^3+b^3}{2(c^3-b^3)}\kappa_0,\label{eq4a}
		\end{equation}
		\begin{equation}
			\kappa_3 =\frac{c^2+b^2}{c^2-b^2}\kappa_0,\label{eq4b}
		\end{equation}
	\end{subequations}
where $c$ is the exradius of the outer layer, and $b$ is the counterpart of the inner layer. Therefore, $\kappa_3$ is determined by $\kappa_0$, which indicates the bilayer thermal cloak will not work when the background changes. In this practical case, we improve the bilayer cloak and overcome its shortcomings by substituting the outer layer with the click metashell, featuring tunable isotropic thermal conductivity that satisfies Eqs.(\ref{eq4a}) and (\ref{eq4b}), as indicated below. Actually, we can adjust the exradius of the inner shell to hold Eqs.(\ref{eq4a}) and (\ref{eq4b}), see Note S3 (Supplementary Materials) \cite{supplemental_material}.
	
\begin{figure}
		\centering
		\includegraphics[width=0.75\linewidth]{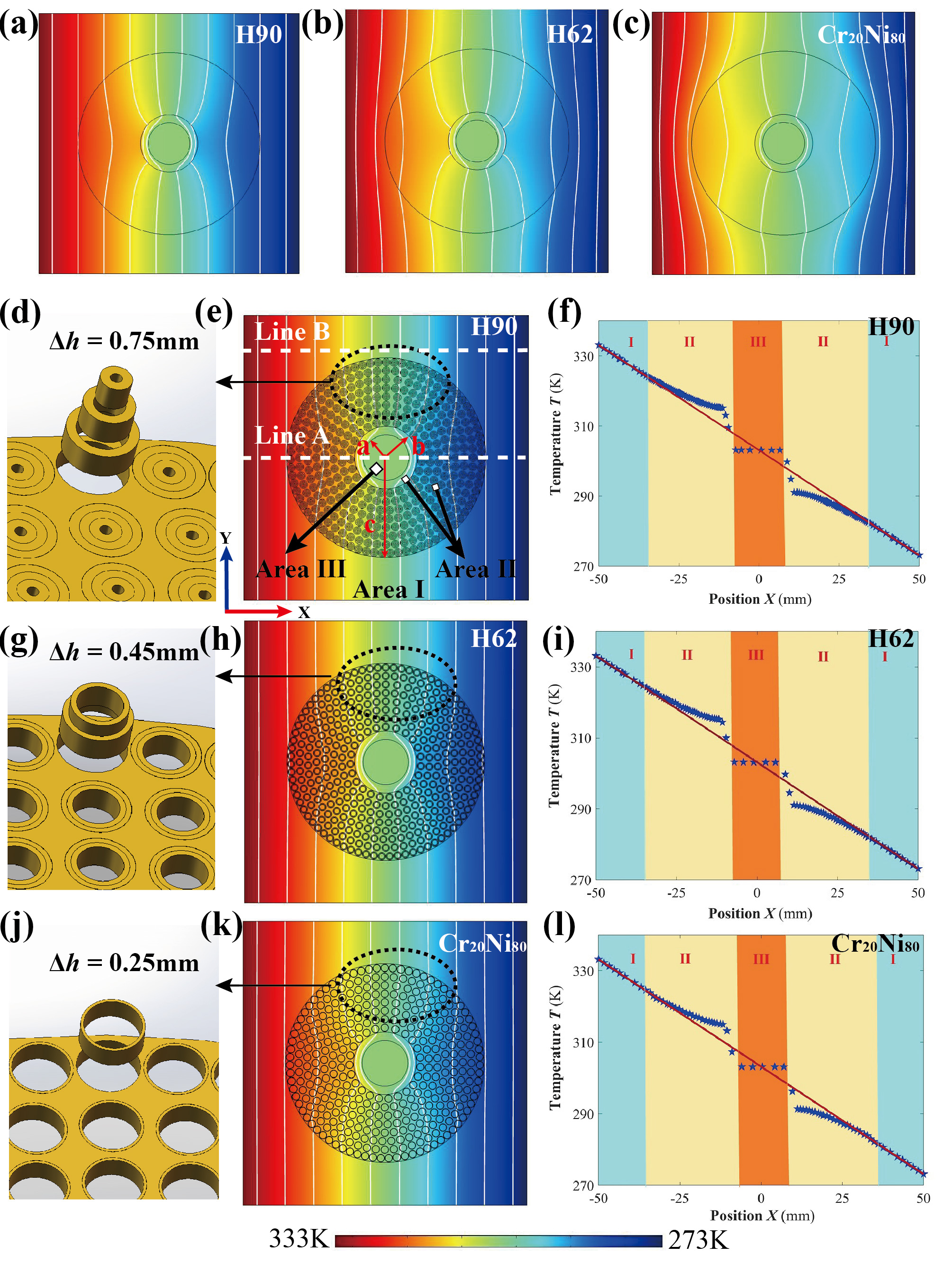}
		\caption{Simulated temperature distributions of the proposed adaptive cloak with $a$=8, $b$=11, $c$=35 mm under three background components. We insulate the upper and lower boundaries and fix the temperature of the left and right borders with 333 K and 273 K. (a)-(c): The bilayer cloak consisting of a fixed outer layer works in three different backgrounds. (d) (g) (j): The THFC's structures in different backgrounds. (e) (h) (k): The simulated temperature distributions of three different backgrounds. The two lines of concern for us are described in (e). (f) (i) (l): The stimulated temperature curves on lines A and B in three different backgrounds. The red line corresponds to the temperature profile along line B, serving as a reference, while the blue stars depict the temperature distribution along line A. The host block (background) is Brass H90 in (e) and (f), Brass H62 in (h) and (i), \protect\chemfig{Cr_{20}Ni_{80}} in (k) and (l).}
		\label{fig.2}
	\end{figure}

Before proceeding with experiments, we perform finite-element simulations to characterize the function of the device. As a contrast, we display the temperature distributions of a traditional bilayer cloak with the fixed outer layer ($\kappa_3$ = 228.7 W m$^{-1}$ K$^{-1}$), working in three different backgrounds; see Figures \ref{fig.2} (a), (b) and (c). Traditional thermal cloak performs perfect thermal cloaking in the Brass (H90), but this effect deteriorates sharply when working in Brass (H62) and \chemfig{Cr_{20}Ni_{80}}. For our adaptive cloak, however, despite being surrounded by three different background components, the temperature distributions outside the cloak (Area I) are not distorted as the external isotherms are straight and equispaced, giving the appearance that nothing is present in the center. And the center areas are isothermal, indicating that the area \uppercase\expandafter{\romannumeral3} is cloaked from the heat flows in three cases; see Figures \ref{fig.2}(e), (h) and (k). Furthermore, to characterize the effectiveness of the thermal cloak quantitatively, the simulated results in Figures \ref{fig.2} (e), (h) and (k) along two horizontal lines [Line A and Line B, depicted in Figures \ref{fig.2}(e)] are shown in Figures \ref{fig.2}(f), (i) and (l). In area \uppercase\expandafter{\romannumeral1} (background), temperature profiles on both two lines are almost identical, illustrating the background temperature distributions are not disturbed. The cloaked object locates in area \uppercase\expandafter{\romannumeral3} where the temperature distributions are uniform. It's obvious that these results are consistently maintained when the background component changes. We also characterize the cloaking performance in the angular direction in Note S2 (Supplementary Materials) \cite{supplemental_material} to verify the adaptive performance.
	
Compared to existing chameleon-like \cite{chameleonlike_metashell_cloak_Xu,chameleonlike_rotator_Yangubao,Chameleonlike_miscrofludics_cloak_Xu} and spinning metamaterials \cite{YingLi_KNI,EMT_Rotating_Qiu,Xu_Tunable} designed to achieve adaptive cloaking, click-reaction-induced paradigm offers a reconfigurable and straightforward preparation process. This overcomes the limitations of previous approaches, where complex principles restricted their practical applications.

\subsection{Convertible Anisotropic Metashell With Functionality Diversities}\label{sec:level3.2}

\begin{figure}
		\centering
		\includegraphics[width=0.8\linewidth]{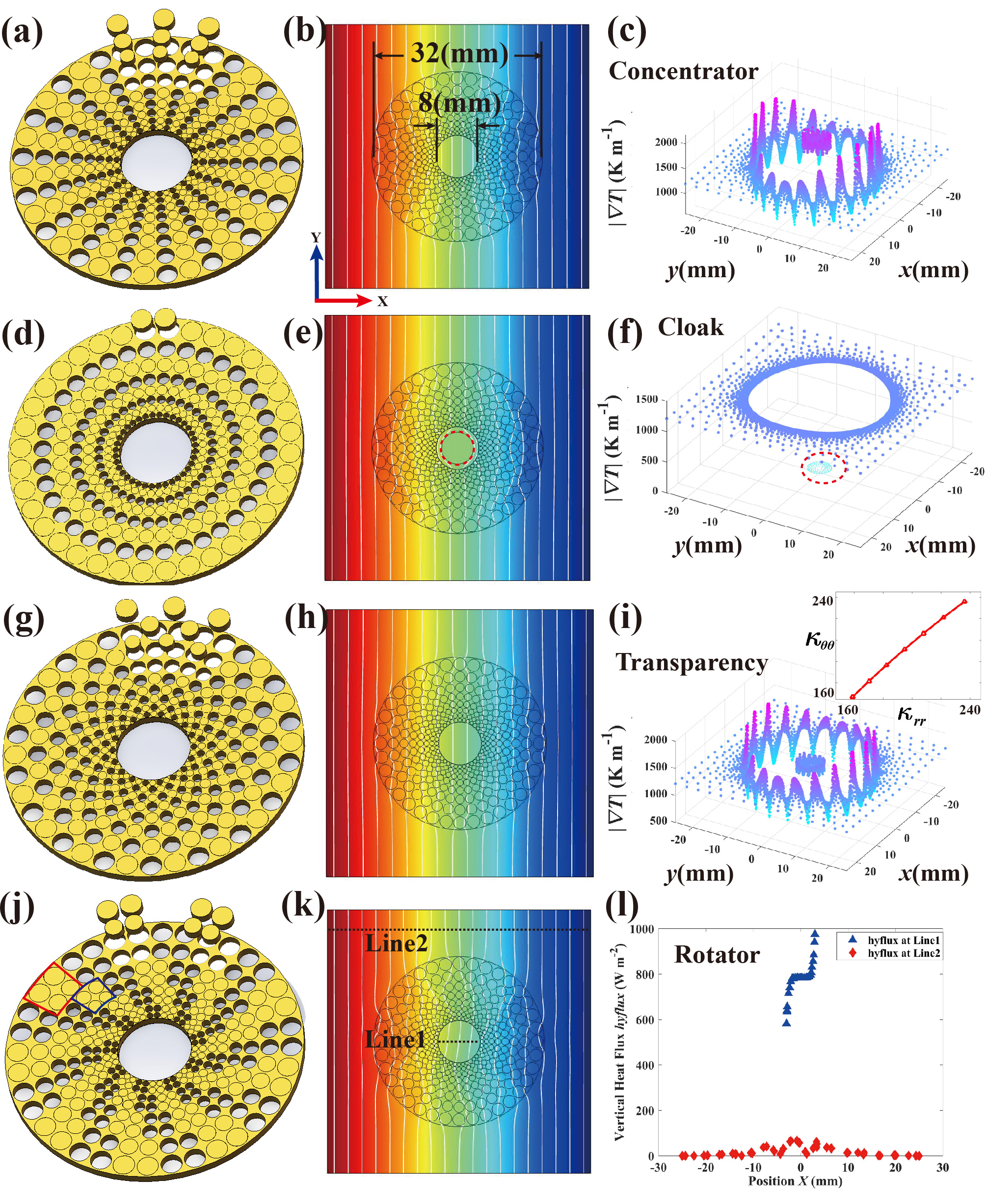}
		\caption{Simulations of the four different functions. (a) (d) (g) (j): Function switching can be realized by altering THFC's state: "hollow" and "filled". (b) (e) (h) (k): The temperature distributions of the thermal concentrator, cloak, transparency device and rotator, respectively. (c) (f) (i): The temperature gradient inside and outside the metashell of the thermal concentrator, cloak and transparency device, respectively. (l): The vertical heat flux of line 1 and line 2.}
		\label{fig.3}
	\end{figure}

Realizing function switching is another significant application of our click metashell, attributable to its flexibility in controlling the degree of thermal anisotropy. For composites whose effective thermal conductivities tensor is $\begin{bmatrix} 
		\kappa_{rr} & \kappa_{r\theta}\\\kappa_{\theta r} & \kappa_{\theta\theta} 
\end{bmatrix}$, the multifunctional device can be designed based on the following principle~\cite{qipan}: (1) When $\kappa_{rr} > \kappa_{\theta\theta}$, the structure serves as a thermal concentrator. (2) When $\kappa_{rr} < \kappa_{\theta\theta}$, the structure serves as a thermal cloak. (3) When $\kappa_{rr} = \kappa_{\theta\theta}$, the structure serves as a thermal transparency device. In this section, we grid the metashell in the same way as described earlier in section \ref{sec:level2.2}. The size of the THFCs varies gradiently along the radial direction while remaining identical along the tangential direction, shown in Figure \ref{fig.1}(a). Keep $\delta \theta = \frac{\pi}{12}$. Our metashell can readily achieve the anisotropic ETC required in different applications, as demonstrated in Figure \ref{fig.1}(c). To create a concentrator, we resort to the orange area of the anisotropic space, see Figure \ref{fig.1}(c). For convenience, we set the air fraction of the UR and LR to zero while simultaneously varying the counterparts of the UL and LL, enabling $\kappa_{rr} > \kappa_{\theta\theta}$. This simplifies the metashell into a binary periodic composite; see Figures \ref{fig.3} (a), (d), (g) and (j). The ETCs of the THFCs with zero air fraction are $\kappa_n$ and that for the changing air fraction is $\kappa_t$. According to Keller’s theory \cite{qipan}, $\kappa_n$ and $\kappa_t$ should satisfy Eq.\ref{eq5} to prevent disturbance of the external temperature field.
	
        \begin{equation}
		\kappa_n\kappa_t=\kappa_0^2
		\label{eq5}
	\end{equation}
where the $\kappa_0$ is the thermal conductivity of the background. In this circumstance, we just need to regulate the ETCs of changing THFCs $\kappa_t$ to comply with Eq.\ref{eq5}. For a thermal cloak, we set the air fraction of the LR and LL to zero but vary the counterparts of the UR and UL simultaneously, which lets the ETCs satisfy $\kappa_{rr} < \kappa_{\theta\theta}$. A thermal insulating layer (expanded polystyrene) is attached between the metashell and the central region, which prevents the heat flow to enter the central region. But such an insulating layer doesn't distort the temperature field of the background due to the effectiveness of the metashell. Although thermal transparency can be realized by modulating the ETC of all the THFCs to $\kappa_0$ simultaneously, we maintain the binary periodic structure because we can readily achieve function switching from the former stage of thermal functions. Meanwhile, the criteria of realizing thermal transparency can be approximately met by setting the air fractions of the diagonal components within the expanded THFCs to zero but varying the anti-diagonal counterparts [see Figure \ref{fig.3}(g)], leading $\kappa_{rr} \approx \kappa_{\theta\theta}$ as depicted in the inset of Figure \ref{fig.3}(i). Furthermore, starting from the outermost side of the shell, we can gradually rotate the expanded THFCs by an angle in a fixed direction as their distance to the center decreases \cite{qipan}, as shown in Figure \ref{fig.3}(j). For instance, the blue expanded THFC can be perceived as the red one rotated by an angle. Such a rotation operation introduces off-diagonal terms in the ETC tensor, as introduced earlier. Consequently, the device is converted into a thermal rotator.
	
Finite-element simulations illustrate the effectiveness of the metashell with different configurations. The periodic arrangement of the expanded THFCs with corresponding structures, as illustrated in Figures \ref{fig.3}(a), (d), (g) and (j), enables the metashell to function as a thermal concentrator, thermal cloak, thermal transparency, and thermal rotator, respectively. The temperature distributions of these four functional devices, under the same temperature setup with Figure \ref{fig.2}, are shown in Figures \ref{fig.3}(b), (e), (h) and (k), respectively, without any distortion in the background. We also calculate the temperature gradient inside and outside the metashell of the thermal concentrator, cloak and transparency, as shown in Figures \ref{fig.3}(c), (f) and (i). In the central region, the concentrator and transparency exhibit larger and similar temperature gradients compared to the background, respectively. In contrast, near-zero values exist in the central region of the thermal cloak. Additionally, we extract the vertical heat flux ($h_y$ flux) along line 1 and line 2; see Figure \ref{fig.3}(l). The appearance of a vertically oriented heat flux in the central region validates the effectiveness of the thermal rotator.

\section{\label{sec:level4}Conclusion and Discussion}

The thermal conductivities of the existing metamaterials are either isotropic or anisotropic, and are invariably fixed. This condition results in a significant shortfall in design flexibility and functional diversity. To overcome these limitations, tunable hollow-filled cells (THFCs) with flexible modification are proposed to design the click metamaterials, corresponding to a counterpart of the modified building blocks in click chemistry. We fabricate a click metashell possessing convertible thermal isotropic or anisotropic conductivity, achieved through the adjustment of THFCs. By discretizing the metashell region into THFCs and controlling their air fraction to adjust the effective thermal conductivity, we achieve full-parameter control over thermal conductivity, enabling convertible thermal conductivities from isotropy to anisotropy and flexible control of thermal anisotropy. The effectiveness of our metashell is demonstrated through its application in an adaptive thermal cloak, capable of responding to varying background components. Meanwhile, it can realize function switching in different working scenes, by simply altering the structures of the expanded THFCs. Both simulations and experiments illustrate that our adaptive cloak can perform exact thermal cloaking under different backgrounds, covering a considerably large range of thermal conductivity. Finite-element simulations illustrate our metashell can switch among the thermal transparency, concentrator, cloak, and rotator. Nearly perfect adaptive (to surroundings, or application requirements) functionalities can be achieved readily without exquisite techniques, indicating our advanced scheme is suitable for practical engineering applications. Therefore, the click metamaterials can realize the fast acquisition of thermal conductivities and functionality diversities. 

Furthermore, it is believed that this scheme could motivate relative research in other fields due to the fundamental technique we used is not limited to a specific physical field. For example, magnetic permeability and electrical conductivity can also be regulated by our method, and a click-reaction-induced adaptive magnetostatic cloak is expected to happen \cite{DCM_bilayer}. Our design undoubtedly provides a practical paradigm about how to readily modify some physical properties of natural materials, which introduces an adjustable degree of freedom in other wave and diffusive systems.

\section{\label{sec:level4}Materials and Methods}

\subsection{Finite-element simulation}

{\bf Adaptive thermal cloak.} We ensure nearly identical shapes for all THFCs, adjusting their air fractions simultaneously using the concentric annuli previously fabricated to modify the ETC. Keep $\delta r_i = r_{i+1} - r_i =$ 3 cm, and the arc length corresponding to the central angle of each THFC is about 3 cm, which enables the THFC is quasi-square. The adaptive bilayer cloak is placed in a host square block (background) with a width of 100 mm and a thickness of 2 mm. Then we insulate the upper and lower boundaries and fix the temperature of the left and right boundaries with 333 K and 273 K. Three different backgrounds we used are Brass (H90) (187.6 W m$^{-1}$ K$^{-1}$), Brass (H62) (116.7 W m$^{-1}$ K$^{-1}$) and \chemfig{Cr_{20}Ni_{80}} (60.3 W m$^{-1}$ K$^{-1}$). The inner and outer layers (metashell) are expanded polystyrene and Aluminum alloy 1060 with the conductivity of 0.02 and 236 W m$^{-1}$ K$^{-1}$, respectively. An Aluminum disk with a radius of 8 mm is placed in the central region as the object that needs to be cloaked. Meanwhile, we choose $b=11$ and $c=35$ mm. Therefore, we get the expected thermal conductivities of the outer layer ($\kappa_3$) using Eq.\ref{eq4b} in three background components; see Table.\ref{table.1}. Then, we obtain the appropriate air fraction of each THFC by resorting to the map datasets in Figure \ref{fig.1}(b); see Table.\ref{table.1}. We fabricate several alternative concentric annuli with thicknesses $\triangle h$ = 0.25 mm, 0.3 mm, 0.35 mm $\cdots$ 0.75mm and 0.8 mm. As shown in Figure \ref{fig.2}(d), after putting three ($\triangle h$ = 0.75 mm) concentric annuli into the THFCs air hole, we can adjust the air fraction to 0.018, thus we ``turn'' the effective thermal conductivity of the metashell (Aluminum alloy 1060; 236 W m$^{-1}$ K$^{-1}$) into about 228 W m$^{-1}$ K$^{-1}$. And supposing we select two $\triangle h$ = 0.45 mm concentric annuli [Figure \ref{fig.2}(g)] or just one $\triangle h$ = 0.25 mm annulus [Figure \ref{fig.2}(j)], we turn the ETC into 142 or 73 W m$^{-1}$ K$^{-1}$, ensuring the adaptive cloak can work under different background components.

\begin{table}[b]
		\caption{The conductivities of three different backgrounds and the corresponding conductivities of the metashell. The last line introduces the air fraction of each THFC in different scenes for achieving adaptive thermal cloaking.}
		\begin{ruledtabular}
        \begin{tabular}{cccc}
			Background&Brass(H90)&Brass(H62)&\chemfig{Cr_{20}Ni_{80}}\\
			\hline
			$\kappa_{0}$(W m$^{-1}$ K$^{-1}$) & 187.6 & 116.7 & 60.3 \\ $\kappa_{3}$(W m$^{-1}$ K$^{-1}$) & 228.7 & 142.28 & 73.52 \\ Air Fraction & 0.018 & 0.25 & 0.52 \\
		\end{tabular}
        \end{ruledtabular}
		\label{table.1}
\end{table}

{\bf Function switching.} The thermal conductivity of the background is 4.24 W m$^{-1}$ K$^{-1}$ ($\kappa_0$), while that of the metashell host is 9 W m$^{-1}$ K$^{-1}$. This indicates that $\kappa_n$ = 9 W m$^{-1}$ K$^{-1}$. For the changing THFCs, we set the air fractions to 0.6, and their ETCs are adjusted to 2.26 W m$^{-1}$ K$^{-1}$ ($\kappa_t$). It's obvious the $\kappa_n$ and $\kappa_t$ satisfy the relation stipulated by Eq.\ref{eq5}. If all air holes in the metashell initially have an air fraction of 0.6, switching between $\kappa_n$ and $\kappa_t$ can be achieved by plugging or unplugging one cylinder of the same size as the air hole. Thus, the air hole only has two states: hollow and filled, which can be easily implemented to realize function switching.

\subsection{Device fabrication and experimental demonstration}

\begin{figure}
		\centering
		\includegraphics[width=1\linewidth]{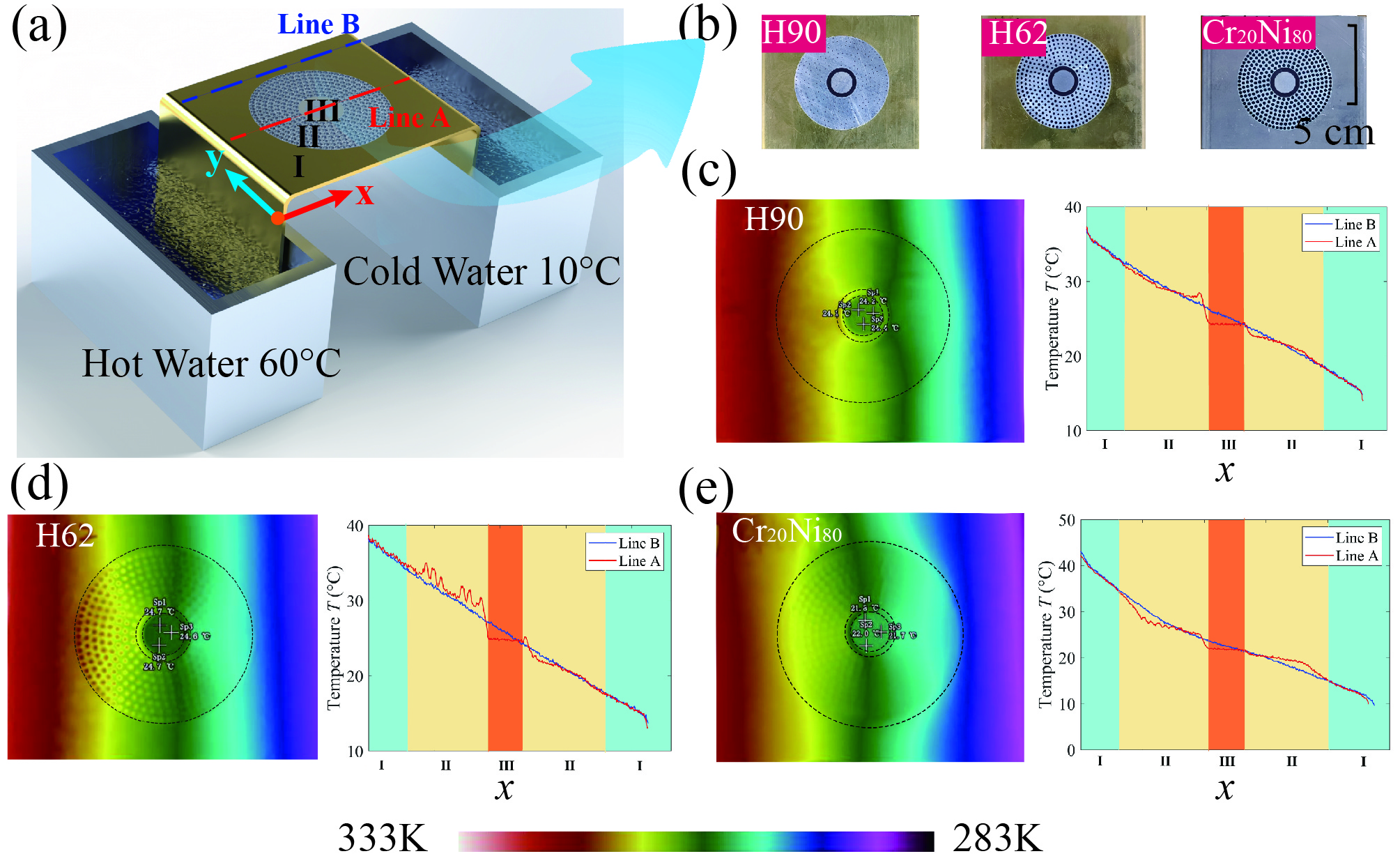}
		\caption{Experimental devices and demonstrations. (a): The experimental device. The left tanker is filled with hot water at \SI{60}{\degreeCelsius} and the right at \SI{10}{\degreeCelsius}. (b): Three specimens with different backgrounds. The experimental results are shown in (c), (d) and (e). The host block (background) is Brass H90 in (c), Brass H62 in (d), \protect\chemfig{Cr_{20}Ni_{80}} in (e).}
		\label{fig.4}
	\end{figure}

For experimental demonstrations, we fabricate a click metashell to form a bilayer thermal cloak. The material of the metashell is Aluminum alloy 1060. Then we put the adaptive thermal cloak into different background components which are Brass (H90), Brass (H62) and \chemfig{Cr_{20}Ni_{80}}, shown in Figure \ref{fig.4}(b). We mechanically embedded the background material with the metashell by extrusion to deduce the contact thermal resistance. The left and right edges of the background plate are bent at right angles for better contact with the source. For convenience, we pre-fabricate three metashells tailored for each of the three operating environments, each possessing the requisite THFCs' (tunable hollow-filled cells) air fraction obtained through direct drilling. The experimental apparatus, schematically illustrated in Figure \ref{fig.4}(a), maintains identical dimensions and components as the simulations. The left and right sides are immersed in containers with hot water at \SI{60}{\degreeCelsius} and cold water at \SI{10}{\degreeCelsius}, respectively. To determine the size of the air holes drilled in the outer layer when the cloak works in a new background, we follow the method proposed in section.\ref{sec:level2.1} and employ the air fraction in Table.\ref{table.1}. We use many ice packs to keep the cold source approximately \SI{10}{\degreeCelsius}, and use a heater to heat the temperature of the heat source to \SI{60}{\degreeCelsius}. The temperature profiles are measured by the infrared camera FLIR E60 when the systems with three background components reach thermodynamic equilibrium, as shown in Figures \ref{fig.4}(c), (d) and (e). The experimental temperature distributions are approximate to the simulated ones shown in Figure \ref{fig.2}, which validate the effectiveness of the adaptive thermal cloak with the click metashell. And we detect the temperature of the Area \uppercase\expandafter{\romannumeral3}. The detection results are about \SI{24.2}{\degreeCelsius}, \SI{24.6}{\degreeCelsius} and \SI{22}{\degreeCelsius} in Figures \ref{fig.4}(c), (d) and (e), respectively. The center areas are always isothermal, regardless of the background component. Both finite-element simulations and experiments prove that the click-reaction-induced thermal cloak can perform adaptive features under different backgrounds.
	
This confirmatory experiment illustrates that adjusting the thermal conductivity of the metashell using THFCs can make it respond adaptively to changes in the environment, namely we can realize convertible thermal isotropy by adjusting the structures of the THFCs. Besides, realizing convertible thermal anisotropy is straightforward as our experiments demonstrate the generality of this approach.

\subsection{Effect of interface thermal resistance}

In relation to the pragmatic applications of thermal metamaterials, a substantial challenge exists in the interfacial thermal resistance encountered between different materials, which frequently leads to temperature discontinuity. Here we discuss the effect of interfacial thermal resistance between concentric annuli aiming to change THFCs' air fractions. Interfacial thermal resistance predominantly comprises two distinct components \cite{ITR_libaowen}: one being the thermal contact resistance arising from suboptimal mechanical contact, and the other being the thermal boundary resistance, resulting from the disparity in physical properties between different materials. The concentric annuli we used are constructed of the same materials as the metashell, whose thermal boundary resistance can be neglected. And we can utilize the forging method of stamping to integrate the concentric annuli, which enhances the contact effectiveness between them, thereby reducing the thermal contact resistance. Moreover, according to both analytical and numerical methodologies \cite{ITR_libaowen,PR_yangshuai}, interfacial thermal resistance plays a significant role at the nanoscale. The interfacial thermal resistance among macro-scale materials used in the construction of multilayer thermal cloaks can be sufficiently minimal as to be disregarded. Given the aforementioned factors, the interfacial thermal resistance can be completely disregarded in this context. On a further note, even if the influence of interfacial thermal resistance is taken into account, it does not diminish the advancement of click metamaterials. This is due to the fact that we would only need to adjust the mapping relationship between the effective thermal conductivity of THFC and the air fraction. Such an adjustment is quite straightforward to compute \cite{ITR_libaowen}.
\\
\\
\noindent{\textbf{Acknowledgments}}

C.W. and P.J. contributed equally to this work. P.J., L.X., and J.H. conceived the idea. C.W. and P.J. designed the model. C.W., P.J., F.Y., L.X., and J.H. discussed the theoretical results and analyzed the model. C.W. and P.J. prepared the samples and performed the experiments. C.W. wrote the original draft. P.J., L.X., and J.H. supervised the research. All authors discussed and contributed to the manuscript. We would like to acknowledge the financial support provided by the National Natural Science Foundation of China under Grant No.~12035004, the Science and Technology Commission of Shanghai Municipality under Grant No.~20JC1414700, and the Innovation Program of Shanghai Municipal Education Commission under Grant No.~2023ZKZD06.
\\
\\
\noindent{\textbf{Data Availability Statement}}

The data that support the findings of this study are available from the corresponding author
upon reasonable request.
\\
\\
\noindent{\textbf{Conflict of Interest}}

The authors declare no conflicts of interest.

\end{document}